\documentclass[aps,prb,reprint,superscriptaddress,longbibliography]{revtex4-2}

\usepackage{graphicx}
\usepackage{dcolumn}
\usepackage{bm}
\usepackage{amsmath}
\usepackage{amssymb}
\usepackage{amsthm}
\usepackage{xcolor}
\usepackage{xspace}
\usepackage{xfrac}
\usepackage{color,soul}
\usepackage{siunitx}
\usepackage{braket}
\usepackage{comment}
\usepackage[colorlinks=true,urlcolor=blue,citecolor=blue,linkcolor=blue,bookmarks=false,pdfstartview={FitH}]{hyperref}
\usepackage{enumitem}
\newcommand{\RR}{\mathbb{R}}
\newcommand{\CC}{\mathbb{C}}

\newcommand{\OO}[1]{\mathcal{O}(#1)}

\newcommand{\abs}[1]{\left| #1 \right|}

\renewcommand{\Im}{\operatorname{Im}}

\newtheorem{thm}{Theorem}

\newcommand{\TBG}{\text{TBG}}

\begin{document}

\author{Jinjing Yi}
\affiliation{Department of Physics and Astronomy, Center for Materials Theory, Rutgers University, Piscataway, NJ 08854, USA}
\affiliation{Center for Computational Quantum Physics, Flatiron Institute, New York, New York 10010, USA}
\author{Daniel Massatt}
\affiliation{Department of Mathematical Sciences, New Jersey Institute of Technology, Newark, NJ, USA}
\author{Andrew Horning}
\affiliation{Department of Mathematical Sciences, Rensselaer Polytechnic Institute, Troy, NY 12180, USA}
\author{Mitchell Luskin}
\affiliation{School of Mathematics, University of Minnesota, Minneapolis, MN 55455 USA}
\author{J. H. Pixley}
\affiliation{Department of Physics and Astronomy, Center for Materials Theory, Rutgers University, Piscataway, NJ 08854, USA}
\affiliation{Center for Computational Quantum Physics, Flatiron Institute, New York, New York 10010, USA}
\author{Jason Kaye}
\affiliation{Center for Computational Quantum Physics, Flatiron Institute, New York, New York 10010, USA}
\affiliation{Center for Computational Mathematics, Flatiron Institute, 162 5th Avenue, New York, NY 10010, USA}

\title{A high-order regularized delta-Chebyshev method for computing spectral densities}

\begin{abstract}
    We introduce a numerical method for computing spectral densities, and apply it to the evaluation of the local density of states (LDOS) of sparse Hamiltonians derived from tight-binding models.
    The approach, which we call the high-order delta-Chebyshev method, can be viewed as a variant of the popular regularized Chebyshev kernel polynomial method (KPM), but it uses a
    high-order accurate approximation of the $\delta$-function to achieve
    rapid convergence to the thermodynamic limit for smooth spectral
    densities. The costly computational steps are identical to those for KPM, with high-order accuracy achieved by an inexpensive post-processing procedure.
    We apply the algorithm
    to tight-binding models of graphene and twisted
    bilayer graphene, demonstrating high-order convergence to the LDOS at non-singular points.
\end{abstract}

\maketitle

\section{Introduction} \label{sec:intro}

A central problem in computational quantum physics is the calculation of spectral functions, which describe the energy spectrum and states of a quantum Hamiltonian operator.
In many applications, one begins with an {\it ab initio} description~\cite{szabo2012modern} of a real material using an effective tight-binding model constructed from Wannier functions \cite{RevModPhys.84.1419,vanderbilt2018berry}, leading to a sparse matrix representation of the Hamiltonian. However, simulating materials in the thermodynamic limit requires increasingly large lattice sizes.
A notable example is moir\'e materials, for which stacking and twisting  
two-dimensional sheets relative to one another produces a moir\'e pattern behaving like an effective lattice with lattice constant on the nanometer (as opposed to angstrom) scale \cite{fang2016,andreitwist,relaxfosdick22,tarnopolsky2019origin,bistritzer2011moire}. Thus, even the large sparse matrix models of these systems can be challenging to handle computationally.

In this setting, direct diagonalization of the Hamiltonian is impossible, and iterative schemes such as the Lanczos method \cite{saadbook} or Chebyshev polynomial-based approaches like the kernel polynomial method (KPM) \cite{KPM,FehskeWeisse2007} are usually favored. These methods aim to approximate the typically continuous spectrum of the thermodynamic limit, and converging efficiently towards this limit is critical. This work proposes a KPM-like algorithm which achieves high-order convergence to the thermodynamic limit away from singular points of the spectrum, improving upon the low-order convergence of the standard KPM. We achieve this by combining recent work on high-order approximations to the $\delta$-function \cite{colbrook21} with a Chebyshev polynomial-based approach to approximating the spectra of tight-binding Hamiltonians.

We focus on the local density of states (LDOS), which can be measured directly in scanning tunneling microscopy experiments~\cite{Tersoff-STM-PRB, RevModPhys.82.1593}. Our approach can also be applied to compute the density of states or the single particle Green's function by supplementing it with an appropriate tracing procedure \cite{KPM,dos17}. 
For a Hamiltonian $\hat H$ on a lattice at position $\bm r$ and energy $E \in \RR$, the LDOS is given by 
\begin{equation} \label{eq:ldos_tl}
    \rho_{\bm r}(E) = \int_{-\infty}^\infty \delta(E - E') \, d\mu_{\bm r}(E')
\end{equation}
(or $d\mu_{\bm r}(E)=\rho_{\bm r}(E)\,dE$),
where $\mu_{\bm r}$ is the spectral measure associated with the state $|\bm r\rangle$. In general, $\rho_{\bm r}(E)$ can be a distribution with both continuous and discrete components, but we assume the typical situation that it is smooth away from a discrete collection of isolated singularities.

For a finite system of dimension $N$ corresponding to a periodic supercell, the spectrum is discrete and $H_N=\sum_{n=1}^N E_n \ket{n}\bra{n}$,
where $E_n$ and $|n\rangle$ denote the eigenenergies and eigenstates of $H_N$, respectively. In this case, the spectral measure is discrete:
\begin{equation}
    \begin{split}
      \mu_{\bm r}^{(N)}([E,E']) &= \sum_{n=1}^N |\langle \bm r| n\rangle|^2 \int_E^{E'} \delta(\lambda - E_n)\;d\lambda\\  
      &=\sum_{E_n\in[E,E']}|\langle \bm r| n\rangle|^2.
    \end{split}
\end{equation}
The LDOS is then given by
\begin{equation} \label{eq:ldos_tlN}\begin{split}
    \rho_{\bm r}^{(N)}(E) &= \int_{-\infty}^\infty \delta(E - E') \, d\mu_{\bm r}^{(N)}(E')\\&=\sum_{n=1}^N|\langle \bm r| n\rangle|^2\delta(E-E_n).
\end{split}
\end{equation}

We say that $\rho^{(N)}_{\bm r} \rightharpoonup\rho_{\bm r}$ (or $\mu^{(N)}_{\bm r} \rightharpoonup \mu_{\bm r}$) in the thermodynamic limit, i.e., as $N \to \infty$,  in the weak sense if \[\int_{-\infty}^\infty \phi(E) \rho^{(N)}_{\bm r}(E)\,dE \to \int_{-\infty}^\infty dE \phi(E) \rho_{\bm r}(E)\,dE\] (or $\int_{-\infty}^\infty \phi(E) \,d\mu^{(N)}_{\bm r}(E) \to \int_{-\infty}^\infty \phi(E)\,d \mu_{\bm r}(E)$\,) for all continuous functions $\phi$. In this case, $\rho^{(N)}_{\bm r}$ can be considered as a distributional approximation of the continuous LDOS $\rho_{\bm r}$ in the thermodynamic limit $N \to \infty$.

Rather than diagonalizing the finite-dimensional approximation $H_N$ of $\hat{H}$ directly, it is typically desirable to obtain a continuous approximation of the thermodynamic LDOS \eqref{eq:ldos_tl} from $H_N$. This can be achieved via a regularization procedure, i.e., by replacing $\rho^{(N)}_{\bm r}(E)$ with
\begin{equation} \label{eq:LDOS_smeared}
    \int_{-\infty}^\infty   k(E, E') \rho^{(N)}_{\bm r}(E')\,dE' = \sum_{n=1}^N |\langle \bm r | n \rangle|^2 k(E, E_n)
\end{equation}
for a suitable regularization kernel $k(E, E')$. For example, Gaussian regularization at scale $\eta$ corresponds to $k(E, E') = \delta_\eta(E - E')$, where $\delta_\eta(E)$ is a Gaussian approximation of the identity with standard deviation $\eta$. In this case, we denote the regularized finite-dimensional LDOS as $\rho_{\bm r}^{(\eta,N)}(E)$. One can also define the regularized LDOS $\rho^{(\eta)}_{\bf r}$ in the thermodynamic limit in an analogous manner, and $\rho^{(\eta)}_{\bf r}(E) \to \rho_{\bf r}(E)$ as $\eta \to 0^+$ for points at which $\rho_{\bf r}(E)$ is continuous. 

\begin{figure}
    \centering

    \includegraphics[width=0.97\linewidth]{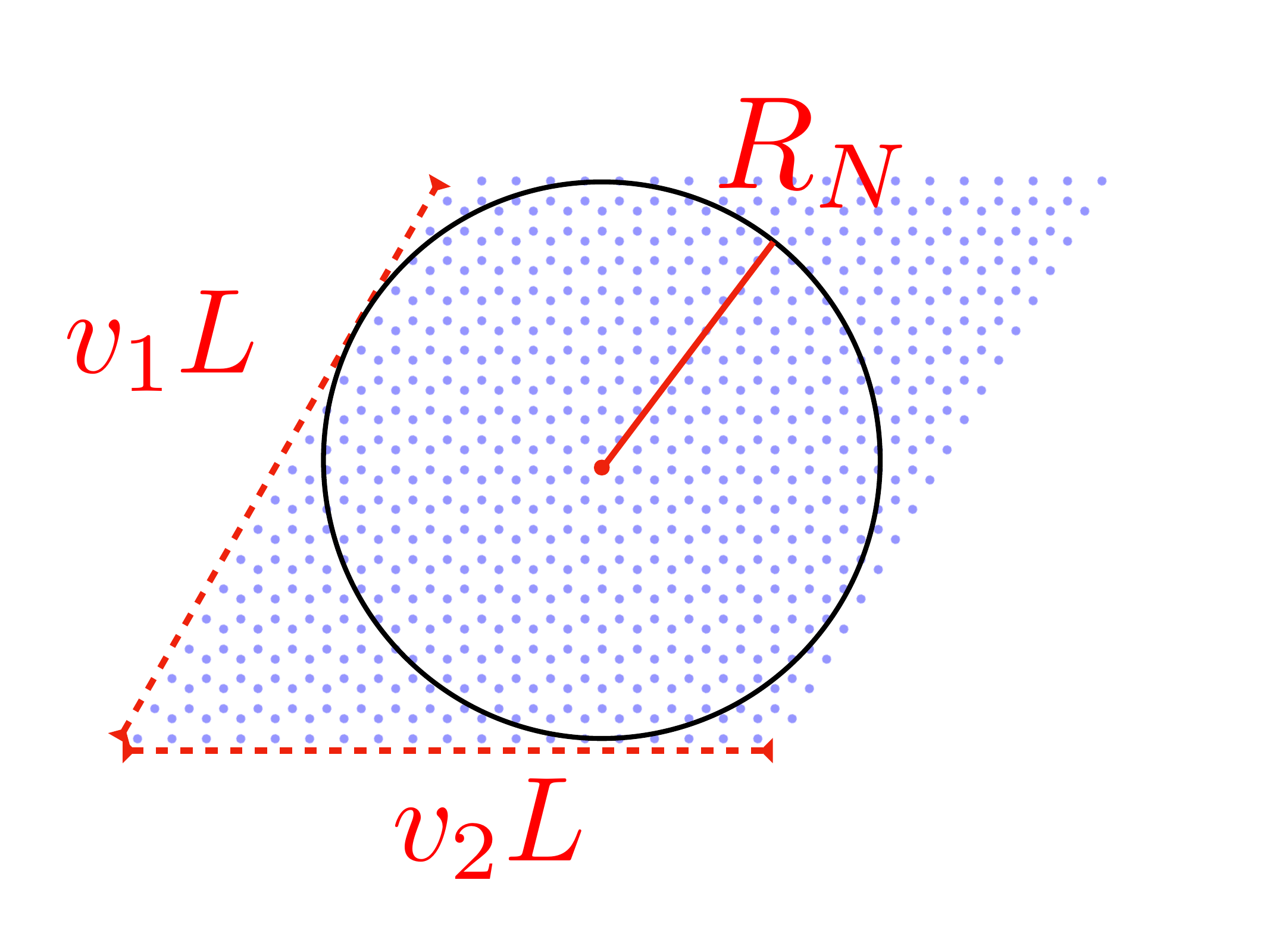} 
    \caption{Monolayer graphene lattice with primitive lattice vectors $v_i$ and supercell lattice vectors $Lv_i$. We illustrate for the highlighted site the corresponding $R_N$, the distance from the site to the edge of the domain.}
    \label{fig:graphene_sc}
\end{figure}

Assuming an exponentially-localized tight-binding model, with $R_N$ the distance between the geometric truncation edge of the finite model $H_N$ and the site $\bm r$ (see Fig.~\ref{fig:graphene_sc}),
we have \cite{dos17,twistronics}
\begin{equation} \label{eq:smearedldosestimate}
    |\rho_{\bm r}^{(\eta)}(E) - \rho_{\bm r}^{(\eta,N)}(E)| \leq C\eta^{-\alpha} e^{-\gamma R_N \eta}
\end{equation}
for some constants $C$, $\alpha$, and $\gamma$ depending on the geometry and hopping functions. We note that while this result is stated for the Gaussian kernel, similar results hold for more general regularization kernels, including the high-order kernels utilized in this work \cite{dos17}. In other words, for a fixed regularization $\eta$, the regularized approximate LDOS converges pointwise to the regularized thermodynamic LDOS as the approximation parameter $N$ is increased.
In order to converge to the thermodynamic LDOS itself, one must take the limits $N \to \infty$ and $\eta \to 0^+$ consistently by making sure $N$ is large enough relative to $\eta^{-1}$. Intuitively, if $N$ is too small relative to $\eta^{-1}$, then the approximation $\rho_{\bm r}^{(\eta,N)}(E)$ will resolve the spurious discrete spectrum of $\rho_{\bm r}^{(N)}(E)$, which is not present in the thermodynamic limit (i.e., in $\rho_{\bm r}(E)$ itself).
In particular, for spectral accuracy with tolerance $\epsilon$, we see from \eqref{eq:smearedldosestimate} that it suffices to choose $R_N \geq -\log(\epsilon\eta^{\alpha'})/(\gamma\eta)$ for $\alpha' > \alpha$.
With this choice, since $\rho_{\bf r}^{(\eta)} \to \rho_{\bf r}$ as $\eta \to 0^+$ at continuous points, $\rho_{\bm r}^{(\eta, N)} \to \rho_{\bf r}$ as well. This procedure thus yields an approximation of the thermodynamic LDOS, itself assumed to be continuous away from isolated points, in terms of regularized, finite-dimensional approximations $\rho_{\bm r}^{(\eta,N)}(E)$. 

Computing the regularized LDOS \eqref{eq:LDOS_smeared} directly requires diagonalizing $H_N$, and is therefore impractical.
KPM \cite{KPM} is a popular approach to
approximating the LDOS and other spectral quantities, which avoids direct diagonalization by using a special choice of the regularization kernel.
In particular, KPM constructs a $p$-term regularized Chebyshev polynomial expansion of the LDOS of $H_N$. 
This can be achieved using only matrix-vector products with $H_N$
by exploiting the Chebyshev three-term recurrence, a process we refer to as
Chebyshev iteration, which is advantageous because $H_N$ is typically a large,
sparse matrix. We discuss the regularization kernel which leads to this scheme in Sec.~\ref{sec:kpm} below.
The expansion size $p$ acts as a regularization parameter, and convergence to the thermodynamic limit is achieved by
increasing $N$ and $p$ simultaneously and consistently \cite{dos17,twistronics}.
Using the method of Jackson regularization, one can achieve $\OO{p^{-2}}$
pointwise convergence to the thermodynamic LDOS away from singular points (see
Sec.~\ref{sec:kpm} and App.~\ref{appendix:jackson}).

In this work, we introduce a method which upgrades this convergence to
$\OO{p^{-m}}$, for an arbitrary fixed positive integer parameter $m$, using the
same primary computational steps as KPM---matrix-vector products with
$H_N$---along with some inexpensive post-processing. The key ingredient is
provided by Ref.~\onlinecite{colbrook21},
which introduces and analyzes a method
of computing a high-order accurate weak approximation of the $\delta$-function as
a sum of simple poles. This achieves $\OO{\eta^m}$ convergence, where $\eta$ is a regularization parameter
characterizing the resolution of the computed spectrum.
That scheme has been used to compute high-order approximations of spectral properties of infinite-dimensional operators in a variety of settings, including spectral measures of differential and integral operators~\cite{colbrook21}, projection-valued measures for tight-binding Hamiltonians of 2D topological insulators~\cite{colbrook2023computing}, non-normalizable states of Schr\"odinger and Poincar\'e differential operators~\cite{colbrook2025computing}, and data-driven spectral decompositions of transfer operators for nonlinear dynamical systems~\cite{colbrook2024rigorous,colbrook2025rigged}.
Our method uses this approximation of the $\delta$-function as a regularization kernel in order to achieve
rapid convergence with respect to $\eta$, and then
approximates the kernel using a Chebyshev expansion to extract the
spectrum of $H_N$ using Chebyshev iteration. 

We make a brief remark contrasting this work with Ref.~\onlinecite{colbrook21}.
There, $\hat{H}$ was taken to be an infinite-dimensional operator, such as a differential or
integro-differential operator, which could be discretized efficiently using
spectral methods to obtain a matrix $H_N$ of moderate size. 
Then, the simple pole approximation of the high-order kernels led to
a collection of linear systems involving $H_N$,
which were solved using direct methods. In our case, we consider 
tight-binding models arising from ab initio approximations of quantum systems,
such that $H_N$ is too large for the direct solution of linear systems. Thus, to
avoid direct linear solves, we obtain Chebyshev polynomial expansions of the
high-order kernels and use Chebyshev iteration to compute the LDOS.

\section{Kernel polynomial method} \label{sec:kpm}

Our algorithm shares many similarities with KPM, which we now review. We refer to Ref.~\onlinecite{KPM} for further details on the method, as well as to Ref.~\onlinecite{linlin16} for a pedagogical introduction to various methods of computing spectral densities, including KPM.
We describe KPM for the LDOS of the discretized Hamiltonian $H_N$, given by \eqref{eq:ldos_tlN}. We assume without loss of generality that the eigenvalues of $H_N$ are contained in $[-1,1]$, otherwise this can be achieved by a shift and rescaling after obtaining a suitable estimate of an interval containing the spectrum of $H_N$.

The objective is to obtain an expansion of \eqref{eq:ldos_tlN} in a Chebyshev series,
\begin{equation} \label{eq:ldoscheb}
\begin{split}
    \rho_{\bm r}^{(N)}(E)  &\equiv \sum_{n=1}^N \abs{ \braket{\bm r | n}}^2 \delta(E - E_n) 
    \\&= \frac{1}{\pi\sqrt{1-E^2}}\sum_{k=0}^\infty \mu_k T_k(E),
    \end{split}
\end{equation}
where the Chebyshev polynomials $T_k(x)$ are defined on the interval $[-1,1]$ and satisfy the orthogonality relations
\begin{equation}
    \int_{-1}^1 dx \, \frac{T_j(x) T_k(x)}{\pi\sqrt{1-x^2}} =  \frac{1+\delta_{0j}}{2} \delta_{jk}.
\end{equation}
Integrating $\rho_{\bm r}^{(N)}(E) $ against $T_k(E)$  therefore gives
\begin{equation} \label{eq:kpmcoefs}
    \begin{aligned}
        \mu_k &= \frac{2}{(1+\delta_{0k})} \sum_{n=1}^N \abs{ \braket{\bm r | n}}^2 T_k(E_n) \\
        &= \frac{2}{(1+\delta_{0k})} \sum_{n=1}^N T_k(E_n) \braket{\bm r | n} \braket{n | \bm r} \\
        &= \frac{2}{(1+\delta_{0k})} \braket{\bm r | {T}_k(H_N) | \bm r},
    \end{aligned}
\end{equation}
where for simplicity we have suppressed the ${\bm r}$-dependence of $\mu_k$.
We note that the single particle Green's function and the global DOS can be obtained in a similar manner~\cite{KPM}.

Using the three-term recurrence
\begin{equation}
    \begin{aligned}
    T_0(x) &= 1, \quad T_1(x) = x, \\
    T_{k+1}(x) &= 2x T_k(x) - T_{k-1}(x),
    \end{aligned}
\end{equation}
the vectors $T_k(H_N) \ket{\bm r}$ can be computed as
\begin{equation} \label{eq:chebrecH}
    \begin{aligned}
        T_0(H_N) \ket{\bm r} &= \ket{\bm r}, \\
        T_1(H_N) \ket{\bm r} &= H_N \ket{\bm r}, \\
        T_{k+1}(H_N) \ket{\bm r} &= 2 H_N T_k(H_N) \ket{\bm r} - T_{k-1}(H_N) \ket{\bm r},
    \end{aligned}
\end{equation}
after which the coefficients $\mu_k$
can be obtained by dot product.
We note that \eqref{eq:chebrecH} only requires matrix-vector products with $H_N$, and storing two vectors of length $N$. The matrix-vector products can often be computed efficiently: for example, if $H_N$ is sparse, one can represent it in a sparse storage format, or apply it on-the-fly.
From the first $p$ Chebyshev coefficients $\mu_k$, we obtain a truncated Chebyshev expansion \eqref{eq:ldoscheb} of $\rho_{\bm r}^{(N)}(E) $ with $k < p$.
If the cost of applying $H_N$ to a vector is $\OO{N}$, as is typical for a sparse representation, then the primary cost of the method is in computing $p-1$ matrix-vector products with $H_N$, giving an $\OO{pN}$ scaling. 

In practice, the approximation of a sum of $\delta$-functions by polynomials leads to Gibbs oscillations, and poor convergence towards the thermodynamic limit. Following \eqref{eq:LDOS_smeared}, this can be addressed by regularizing \eqref{eq:ldoscheb}: 
\begin{equation} \label{eq:ldoscheb2}
\begin{split}
  \rho_{\bm r}^{(N)}(E)\approx   \rho_{\bm r}^{(N,p)}(E)  &\equiv \sum_{n=1}^N \abs{ \braket{\bm r | n}}^2 k_p(E,E_n) 
    \\&= \frac{1}{\pi\sqrt{1-E^2}}\sum_{k=0}^\infty \nu_k T_k(E).
    \end{split}
\end{equation}
KPM typically uses a regularization kernel
\begin{equation}
\label{eq:kpm_kernel}
k_p(E,E') = \frac{1}{\pi\sqrt{1-E^2}}\sum_{k = 0}^{p-1} \frac{2 g_k^p}{1+\delta_{0k}} T_k(E)T_k(E'),
\end{equation}
with weights $g_k^p$, to be discussed below, yielding $\delta(E-E')$ distributionally in the limit $p \to \infty$. Repeating the procedure above yields the regularized weights
\begin{equation} \label{eq:regwgts}
    \nu_k = \mu_k g_k^p
\end{equation}
for $k < p$, and zero otherwise.
Thus the regularized procedure can be implemented by following precisely the same steps as above, and multiplying the coefficients $\mu_k$ by regularization weights. This is the motivation for choosing the regularization kernel in the form \eqref{eq:kpm_kernel}.

We note that this regularized version of the method, rather than the unregularized formulation presented above, is often referred to as KPM.
For $g_n^p = 1$, i.e., no regularization, $k_p$ is the Dirichlet kernel. The so-called Jackson kernel selects $g_n^p$ to preserve positivity of the kernel (and consequently the density), minimize variance, and preserve the total density. The Jackson coefficients are given by
\begin{equation} \label{eq:jackson_wgts}
    g_k^p = \frac{(p+1-k)\cos\frac{\pi k}{p+1} + \sin \frac{\pi k}{p+1} \cot \frac{\pi}{p+1}}{p+1}.
\end{equation}

\begin{figure}
    \centering
    \includegraphics[width=\linewidth]{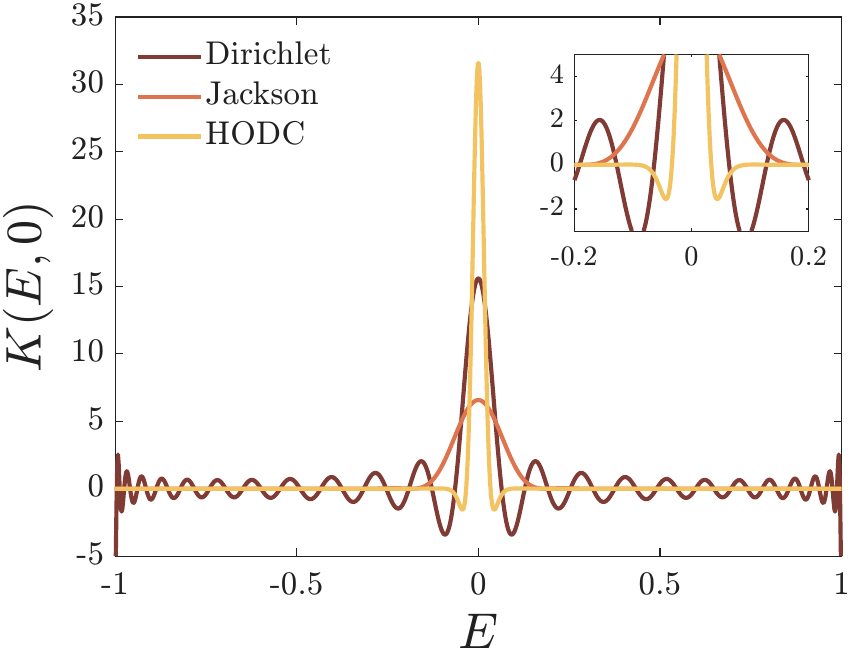} 
    \caption{Dirichlet and Jackson kernels $k_p(E, 0)$ with $p = 50$, and high-order kernel $K_\eta(E,0)$ with $m = 6$ and $\eta = 0.05$.}
    \label{fig:kernels}
\end{figure}

The width of the Jackson kernel scales as $1/p$, which at sufficiently smooth points
of the density of interest (in the thermodynamic limit) guarantees $\OO{p^{-2}}$ convergence of the KPM approximation (see App.~\ref{appendix:jackson}). Fig.~\ref{fig:kernels} shows a comparison of the Dirichlet kernel, which exhibits Gibbs oscillations, and the Jackson kernel, which is smooth and positive.

\section{High-order regularized Delta-Chebyshev method} \label{sec:hodc}

To improve upon the convergence rate of KPM with Jackson regularization, we take the closely related Delta-Gauss-Legendre (DGL) method \cite{linlin16} as a starting point. The DGL method proceeds as follows: (1) the $\delta$-function in \eqref{eq:ldos_tlN} is replaced by a Gaussian of standard deviation $\eta$, (2) the Gaussian is approximated by a Legendre polynomial expansion, and (3) the Hamiltonian $H_N$ is substituted into the resulting expression, and evaluated using the three-term recurrence for Legendre polynomials. We note that the use of Legendre rather than Chebyshev polynomials in this procedure is simply a convenience, since in that case in that case analytical expressions for the Gaussian expansion coefficients are available. The bottleneck computational step in the resulting method is the same as for KPM: the evaluation of a three-term recurrence. Compared with KPM, this method introduces an additional regularization parameter $\eta$ which must be taken to zero as $N$ is increased to obtain the LDOS in the thermodynamic limit, in the manner discussed in Sec.~\ref{sec:intro}.

By replacing the Gaussian in the DGL method by the high-order kernels introduced in Ref.~\onlinecite{colbrook21}, we can obtain a similar method which yields high-order convergence to the LDOS in the thermodynamic limit at a negligible additional cost. 
It can be shown that a Gaussian is only a second-order kernel in this sense, so that the DGL method alone is insufficient to obtain high-order convergence to the thermodynamic limit.

App.~\ref{sec:appendix_HOK} contains a brief introduction to the rational high-order kernels we use here, including their construction and approximation properties. An $m$th-order rational high-order kernel is an approximation of the shifted $\delta$-function $\delta(E - x)$ of the following form \cite{colbrook21}:
\begin{equation}\label{eq:Keta}
   K_\eta(E,x) = -\frac{1}{\pi}\sum_{l=1}^m \Im \frac{w_l}{E - x + \eta z_l}.
\end{equation}
Here the weights $w_l \in \CC$ and pole locations $z_l \in \CC$ are chosen such that $K_\eta(E,x) \to
\delta(E-x)$ weakly as $\OO{\eta^m}$. 
There are various ways of choosing the pole
locations. One is to take
\begin{equation}\label{eqn:kernel_poles}
z_l = x_l + i,
\end{equation}
for $x_l \in \RR$, and to choose the $w_l$ according to a moment-matching condition which yields the desired order of
accuracy. The resulting weights and pole locations are universal, i.e., independent of the Hamiltonian, and can be pre-tabulated. We refer to App.~\ref{sec:appendix_HOK} and Ref.~\onlinecite{colbrook21} for further details on the method of calculating the poles and weights, and analysis of the kernels. 
The kernel for $m=6$ and $\eta = 0.05$ is shown in Fig.~\ref{fig:kernels}, along with the Dirichlet and Jackson kernels for comparison.

We again assume, after a shift and rescaling, that the spectrum of the approximate Hamiltonian $H_N$ is contained in the interval $[-1, 1]$. 
Since for each $E \in \RR$, $K_\eta(E,x)$ is the imaginary part of a function which can be extended analytically to the strip $\abs{\Im x} < \eta$, we can approximate it by a Chebyshev expansion,
\begin{equation}\label{eq:Ketacheb}
    K_\eta(E,x) \approx \sum_{k=0}^{p-1} \nu_k(E, \eta) {T}_k(x),
\end{equation}
with exponential convergence in $p$ \cite{trefethen19}. We note that the kernel is not written in the KPM form \eqref{eq:kpm_kernel}, so we do not obtain regularization weights of the form \eqref{eq:regwgts}.
From the analyticity in the strip, it can be shown from standard estimates that the error of such an expansion decreases with $p$ exponentially with rate proportional to $\eta$ \cite{trefethen19}, so that $p = \OO{1/\eta}$ is required to achieve a fixed accuracy. Given $E$ and $\eta$, the Chebyshev coefficients $\nu_k(E, \eta)$ can be computed efficiently by evaluating \eqref{eq:Keta} at Chebyshev nodes and applying a fast cosine transform, at an $\OO{p \log p}$ cost \cite{trefethen19}. 

We next consider the LDOS of $H_N$ regularized by the kernel $K_\eta$: 
\begin{equation} \label{eq:hodcderivation}
    \begin{aligned}
        \rho_{\bm r}^{(\eta, N)}(E) &= \sum_{n=1}^N \abs{\braket{\bm r | n}}^2 K_\eta(E, E_n) \\
        &\approx \sum_{n=1}^N \abs{\braket{\bm r | n}}^2 \sum_{k=0}^{p-1} \nu_k(E, \eta) T_k(E_n) \\
        &= \sum_{k=0}^{p-1} \nu_k(E,\eta) \bra{\bm r}{T}_k(H_N) \ket{\bm r}.
    \end{aligned}
\end{equation}
Here, the last line is obtained in a manner similar to \eqref{eq:kpmcoefs}.
As in KPM, the primary computational cost is computing the Chebyshev moments $\bra{\bm r} {T}_k(H_N) \ket{\bm r}$, which can again be done using the three-term recurrence \eqref{eq:chebrecH}. Once the moments have been obtained, evaluating the approximation of $\rho_{\bm r}^{(\eta,N)}(E)$ at a particular choice of $E$ is a straightforward and inexpensive post-processing step: we obtain the Chebyshev coefficients $\nu_k(E,\eta)$ of $K_\eta(E,x)$ for $k=0,\ldots,p-1$ using a fast cosine transform, and then evaluate the final expression in \eqref{eq:hodcderivation}. The use of an $m$th-order kernel yields a convergence rate $\OO{\eta^m} = \OO{p^{-m}}$ to the thermodynamic limit $\rho_{\bm r}(E)$ away from singular points, rather than the $\OO{p^{-2}}$ rate of KPM with Jackson smoothing.
The additional cost associated with evaluating $\rho_{\bm r}^{(\eta, N)}(E)$ given the Chebyshev moments is insignificant compared with the cost of computing the moments themselves, so the improved convergence rate is achieved at a negligible additional cost compared with KPM. 

We note two possible disadvantages compared with the Jackson KPM method. First, unlike the Jackson kernel, the high-order kernel is not strictly positive (see Fig.~\ref{fig:kernels}), so the resulting LDOS will not in general be strictly positive either. Indeed, it is not possible to achieve high-order convergence in this manner using a strictly positive kernel \cite{KPM}. 
However, since the convergence rate can be made significantly faster, we expect an overall smaller error.

Second, the method requires carefully choosing the regularization parameter $\eta$ in addition to the Chebyshev truncation parameter $p$, whereas KPM only requires choosing $p$. We make $\eta$ the independent parameter, and automate the selection of $p$ as follows. As discussed above, since $K_\eta(E,x)$ is smooth, the approximation \eqref{eq:Ketacheb} converges exponentially with respect to $p$, and therefore so does the approximation $\eqref{eq:hodcderivation}$ of $\rho_{\bm r}^{(\eta,N)}(E)$ for fixed $\eta$ and $N$. We therefore increase $p$ in steps, adding terms to \eqref{eq:hodcderivation} until the difference in the result for all $E$ falls below a specified tolerance $\epsilon$. This ensures a nearly optimal choice of $p$ for each $\eta$.

\begin{figure}
    \centering
    \includegraphics[width=0.97\linewidth]{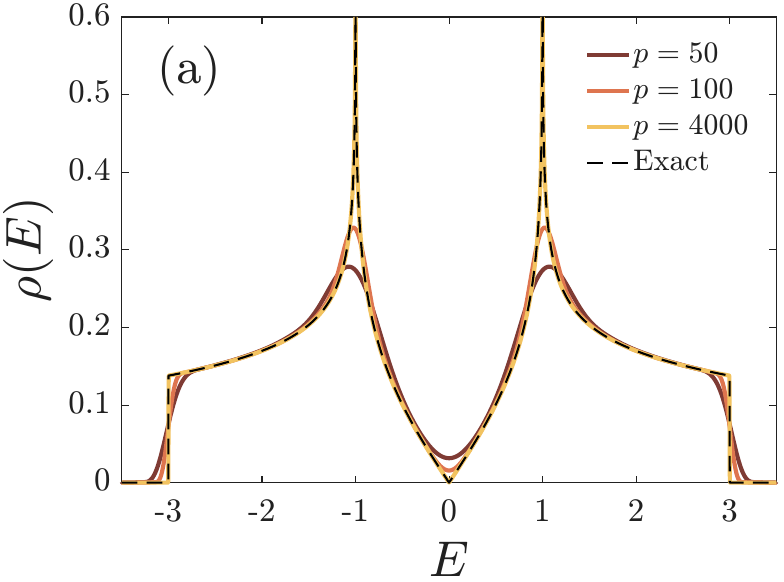}
    \hfill
    \includegraphics[width=0.97\linewidth]{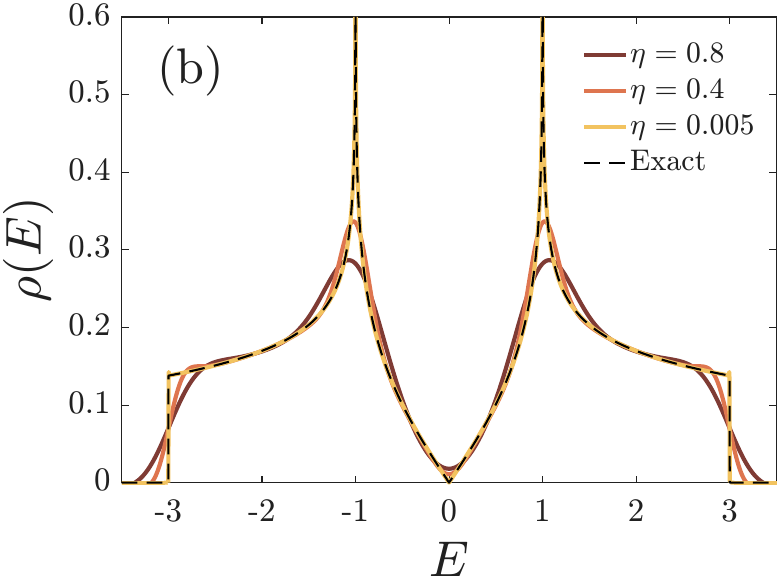} 
    \caption{Density of states for nearest-neighbor tight-binding model of graphene produced by (a) Jackson KPM and (b) the HODC method, for several values of (a) the Chebyshev degree parameter $p$ and (b) the regularization parameter $\eta$. Dashed black line shows the exact density of states.}
    \label{fig:graphene_dos}
\end{figure}

\section{Numerical examples}

We consider two applications of our approach. For the first, single-layer graphene, the LDOS is known analytically, and can be used to test the accuracy of our scheme. The second is twisted bilayer graphene (TBG) near the magic angle of $\theta \approx 1.1^{\circ}$~\cite{bistritzer2011moire}. For this system, the LDOS is not known analytically, and a large lattice is required to capture the low-energy effective bands representing tunneling on the moir\'e scale (see Fig.~\ref{fig:disk}).

\subsection{Graphene}

The second-quantized Hamiltonian of single-layer graphene, taking only one of the degenerate spin components, is given by
\begin{equation}
    H_G = -t\sum_{\langle i,j\rangle} (a^{\dagger}_{i}a_{j}+h.c.),
    \label{eqn:HG}
\end{equation}
where $t$ is a hopping parameter, $a_i^\dagger$ ($a_i$) is the creation (annihilation) operator at site $i$ on the honeycomb lattice, 
and $\sum_{\langle i,j\rangle}$ indicates a lattice sum over three nearest neighbors.
The LDOS is given analytically by \cite{graphene},
\begin{equation}
\rho(E) = 
\begin{cases}
    \frac{|E|}{\pi^2 t^2 \sqrt{Z_0(E)}} K\left(\frac{Z_1(E)}{Z_0(E)}\right) , & |E| \leq 3t, \\
    0, & \text{otherwise},
\end{cases}
\end{equation}
where $K$ is the complete elliptic integral of the first kind, and 
\begin{equation}
\begin{aligned}
Z_0(E) &= \begin{cases}\left(1+\left|\frac{E}{t}\right|\right)^2-\frac{\left[(E / t)^2-1\right]^2}{4}, & \abs{E} \leq t, \\
4\left|\frac{E}{t}\right|, & t < \abs{E} \leq 3t,\end{cases} \\
Z_1(E) &= \begin{cases}4\left|\frac{E}{t}\right|, & \abs{E} \leq t ,\\
\left(1+\left|\frac{E}{t}\right|\right)^2-\frac{\left[(E / t)^2-1\right]^2}{4}, & t < \abs{E} \leq 3t.\end{cases}
\end{aligned}
\end{equation}
We note that the LDOS is site-independent in this case, so that the global DOS is identical to the LDOS up to a rescaling.
To truncate the system, we select a supercell containing all sites within $(v_1,v_2)[0,L),$
where $v_1$ and $v_2$ are the lattice vectors. Then we apply periodic boundary conditions to the tight-binding Hamiltonian.

\begin{figure}
    \centering
\includegraphics[width=0.96\linewidth]{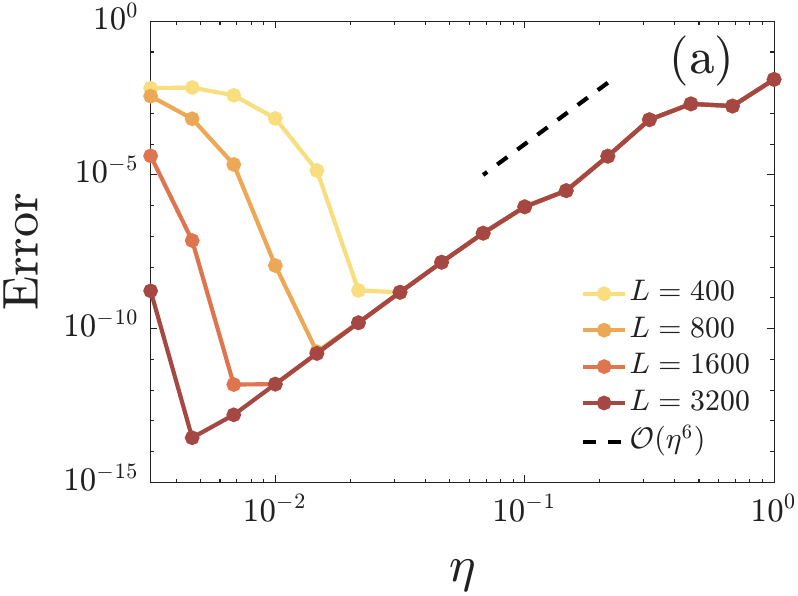}
    \hfill
    \includegraphics[width=0.96\linewidth]{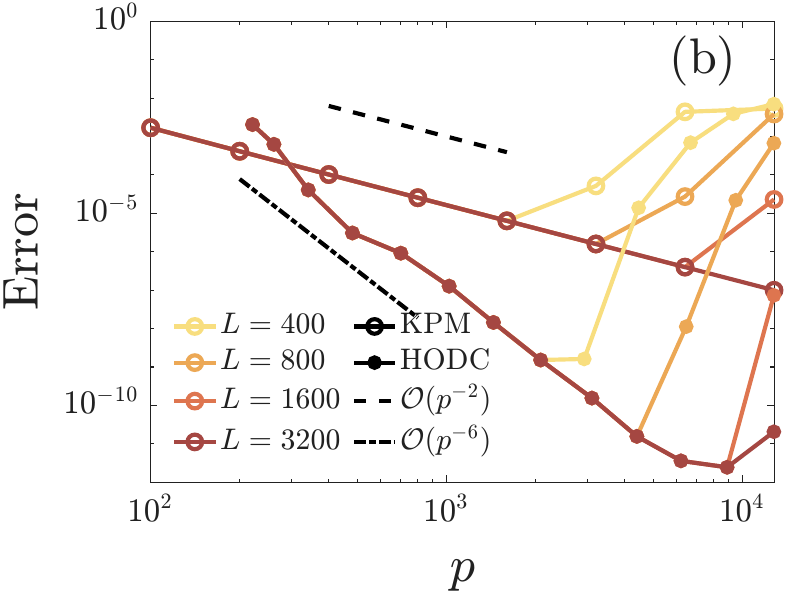}
    \caption{Error of the HODC scheme for the graphene example at $E = 0.5$. (Top) $\OO{\eta^m}$ convergence for $m = 6$, using fixed, sufficiently large Chebyshev degree $p$. For sufficiently small $\eta$, finite size effects amplify the error, requiring larger system sizes $L$. (Bottom) Convergence of Jackson KPM and HODC with $m = 6$. For HODC, $\eta$ is varied, and the Chebyshev order $p$ is adaptively selected based on an error tolerance $\epsilon = 10^{-12}$.}
    \label{fig:graphene_dos_hodc_etaconverge}
\end{figure}

The exact LDOS is shown in Fig.~\ref{fig:graphene_dos}, along with its approximation using KPM with Jackson regularization for several choices of the Chebyshev truncation parameter $p$ (panel (a)), and using the HODC method with order $m = 6$ for several choices of the regularization parameter $\eta$ (panel (b)). In both cases we use the lattice truncation parameter $L = 1600$, which we have determined is large enough to avoid significant finite-size effects. Both methods show visual convergence to the thermodynamic limit $\rho$ as the regularization parameter is decreased.

In Fig.~\ref{fig:graphene_dos_hodc_etaconverge}, we examine pointwise convergence to the thermodynamic limit at $E = 0.5$, which is well-separated from singularities. For the HODC method, we first fix a large value of $p$, and vary $\eta$ for different system sizes $L$ (panel (a)). For $m = 6$, we observe the expected $\OO{\eta^6}$ convergence until finite-size effects begin to dominate: the method begins to resolve the discrete spectrum of the finite-dimensional Hamiltonian, and the error increases. Increasing $L$ pushes this effect to smaller values of $\eta$, allowing for smaller minimum errors with respect to the thermodynamic limit. This convergence behavior is consistent with the discussion in Sec.~\ref{sec:intro}. Next, using the procedure described at the end of Sec.~\ref{sec:hodc} to determine $p$ from $\eta$, we compare the convergence of KPM with Jackson smoothing and HODC with respect to $p$ (panel (b)). We observe the expected convergence rates (as well as the same finite-size effect as before), with HODC achieving better error than KPM beyond $p \approx 200$. We note that the particular pointwise convergence behavior is dependent on $E$, and the high-order convergence of HODC vanishes in the vicinity of singularities (here this corresponds to van-Hove singularities and the Dirac point).
\begin{figure}
    \centering
    \includegraphics[width=0.97\linewidth]{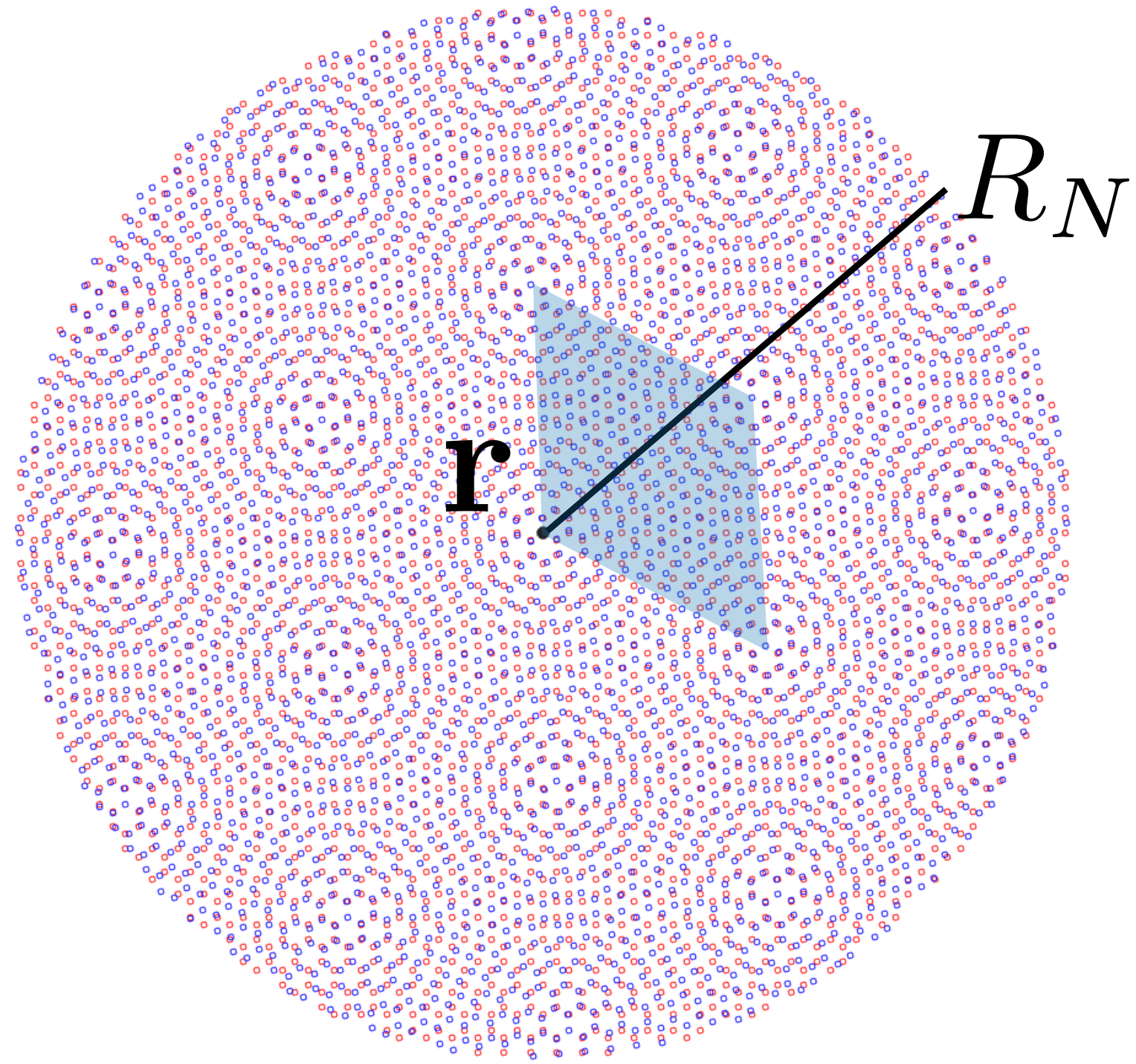} 
    \caption{A cut-out of TBG of radius $R_N$ centered at the site ${\bf r}$. The shaded parallelogram corresponds to a moiré unit cell.} 
    \label{fig:disk}
\end{figure}

\begin{figure*}
\includegraphics[width=\linewidth]{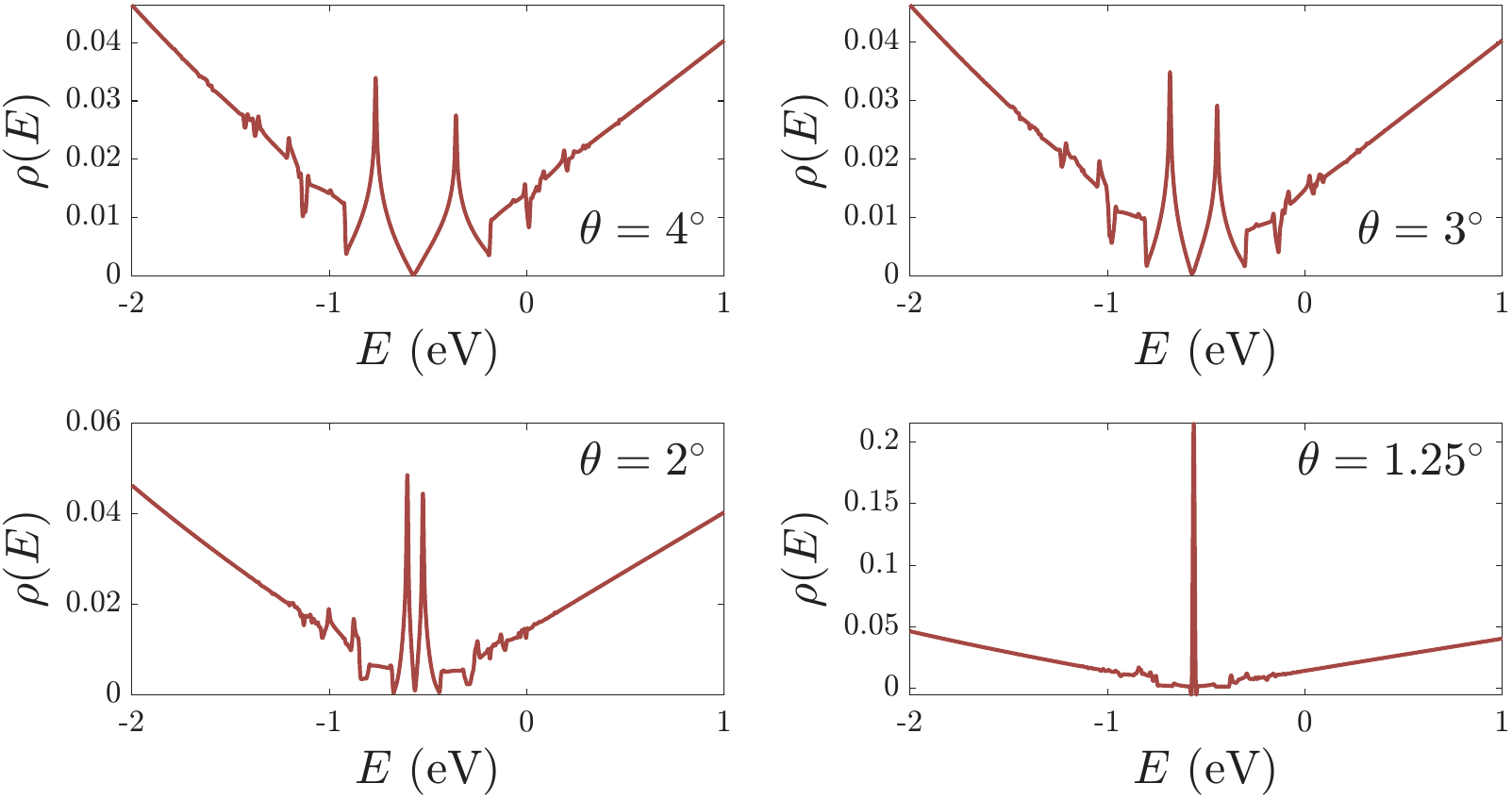}
    \caption{Local density of states for twisted bilayer graphene model on an energy window near the Fermi level, produced using HODC with $m = 6$, for several values of the twist angle $\theta$.
    }
    \label{fig:tbg_ldos_manytheta}
\end{figure*}

\subsection{Twisted Bilayer Graphene}

We next consider twisted bilayer graphene (TBG), obtained by stacking two single layers of graphene, offset with a relative rotation by an angle $\theta$. The close proximity of the two layers leads to electronic tunneling between them (approximately $0.11$ meV), making TBG distinct from single-layer graphene. Many numerical methods have been developed to model the spectrum of TBG,
at different levels of physical approximation~\cite{bistritzer2011moire,PhysRevLett.99.256802,twistronics, tarnopolsky2019origin, fu2020magic,PhysRevResearch.2.023325, PhysRevB.101.235121, PhysRevB.103.205412,jauslin2025incommensurate}. It is not possible to handle the corresponding tight binding problem exactly while retaining exponential convergence to the thermodynamic limit~\cite{PhysRevB.98.224102,PhysRevB.86.155449,fu2020magic,PhysRevB.103.205412}. In the following, we therefore turn to the approximation based on the ``local configuration'' method~\cite{genkubo17,twistronics,dos17}. This treats the limit of weak incommensurate effects (applicable to TBG near the magic angle) to good accuracy while allowing for exponential convergence to the thermodynamic limit.

We use the Wannier tight-binding model and parameters developed in Ref.~\onlinecite{fang2016}. For each layer of graphene, this model includes up to four nearest neighbor couplings (6 \AA), unlike the nearest-neighbor model of Eq.~\eqref{eqn:HG}. 
The intralayer graphene Hamiltonian of layer $\ell$ is therefore given by
\begin{equation}
    H_\ell= \sum_{ i,j}( t_{ij} a^\dagger_{i,\ell} a_{j,\ell}+\mathrm{h.c.}),
\end{equation}
with $i,j$ the atomic positions 
and the nearest neighbor intralayer tunneling energies given by \cite{fang2016}
\begin{equation*}t_1=-2.8922,\ t_2=0.2425,\  t_3=-0.2656,\  t_4=0.0235, 
\end{equation*}
where $t_{ij}=t_m$ if the $i,j$ atomic positions are $m$th nearest neighbors \cite{jungmacdonaldtbmodel}. The on-site energies are given by $t_{i,i}=\epsilon_C=0.3504$. 

\begin{figure*}
\includegraphics[width=\linewidth]{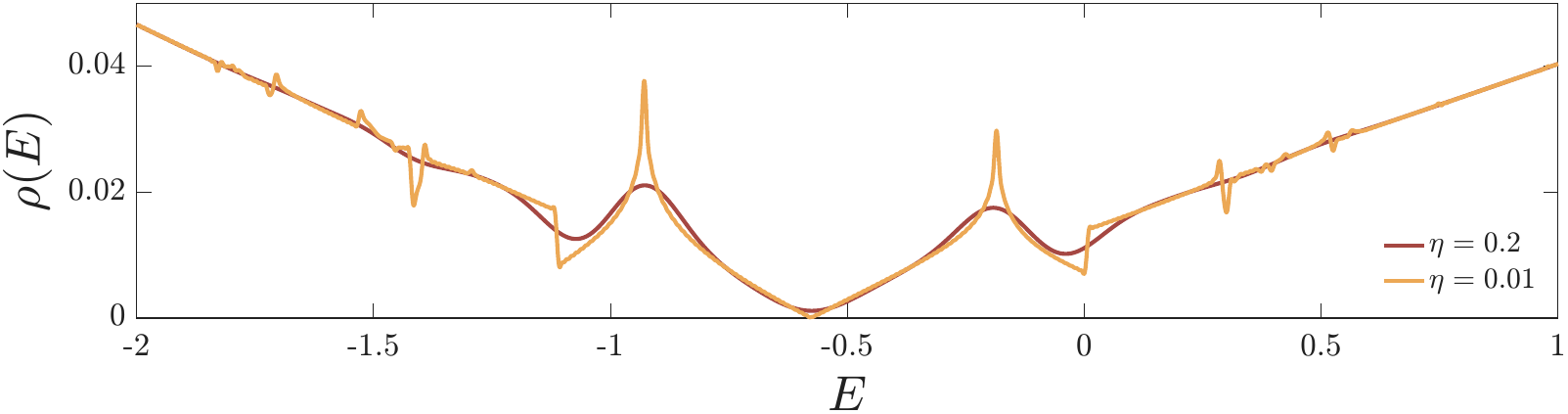} 
\caption{Local density of states for twisted bilayer graphene model with $\theta = 6^\circ$, on an energy window near the Fermi level. The LDOS was produced using the HODC method with $m = 6$, and two different choices of $\eta$.}
\label{fig:tbg_ldos}
\end{figure*}

The interlayer Hamiltonian is given by
\begin{equation} \label{eq:Hinter}
    H_{\text{inter}} = \sum_{i,j} t_\theta(\vec{r}_i-\vec{r}_j)a_{i,1}^\dagger a_{j,2} + \text{h.c.}
\end{equation}
for lattice site $i$ at position $\vec{r}_i$ in layer 1 and site $j$ at position $\vec{r}_j$ in layer 2.
Here the twisted interlayer tunneling $t_\theta({\bf r})$ is an exponentially decaying function fitted to the tunneling between Wannier orbitals of adjacent layers through the sum of the monolayer DFT potentials. 
The hopping energy is approximated in Ref.~\onlinecite{fang2016} and we include it here for completeness:
\begin{equation}
\begin{split}
t_\theta(\vec{r})= & V_0(r) +V_3(r) (\cos(3\theta_{\rm 12})+\cos(3\theta_{\rm 21})) \\
& + V_6(r) (\cos(6\theta_{\rm 12})+\cos(6\theta_{\rm 21})).
\end{split}
\label{eqn:TBH_interlayer}
\end{equation}
Here $\vec{r}=\vec{r}_i-\vec{r}_j$ is the two-dimensional (projected) vector connecting the two atoms, $r=|\vec{r}|$, and $\theta_{12}$, $\theta_{21}$ are the angles between the projected interlayer bond and the in-plane nearest neighbor bond (see Fig.~1 in Ref.~\onlinecite{fang2016}).
The fitting functions $V_i(r)$ are given by  
\begin{equation}
\begin{split}
V_0(r) & =  \lambda_0 e^{-\xi_0 \bar{r}^2} \cos(\kappa_0 \bar{r}) \\
V_3(r) & =  \lambda_3 \bar{r}^2 e^{-\xi_3(\bar{r}-x_3)^2} \\
V_6(r) & =  \lambda_6 e^{-\xi_6(\bar{r}-x_6)^2} \sin(\kappa_6 \bar{r})
\end{split}
\label{eqn:TBH_interlayer_fit}
\end{equation}
with $\bar{r}=r/a$, $a=2.46$ \AA, and parameters given by
\begin{center}
\begin{tabular}{|c c c c|} 
 \hline
 $i=$& \textbf{$0$} & \textbf{$3$}  & \textbf{$6$}  \\
 \hline
 \textbf{$\lambda_i$} & $0.3155$ & $-0.0688$ & $-0.0083$ \\ 
 \hline
 \textbf{$\xi_i$} &$1.7543$          & $3.4692$          & $2.8764$ \\
 \hline
 \textbf{$x_i$}  & $-$               & $0.5212$          & $1.5206$\\
 \hline
 \textbf{$\kappa_i$} & $2.0010$          & $-$               & $1.5731$ \\
 \hline
\end{tabular}.
\end{center}
The full TBG Hamiltonian is then 
\begin{equation}
    H_\TBG = H_1 + H_2 + H_\text{inter}.
\end{equation}

The DOS and LDOS of TBG are far more sophisticated than those of monolayer graphene. First, the twist downfolds the low energy bands into a mini Brillouin zone. Second, as shown in Fig.~\ref{fig:disk}, the superlattice geometry enlarges the approximate translational symmetry from the angstrom to the nanometer scale. As shown there, we truncate the system using a radial truncation of radius $R_N$ centered at an AA-stacked lattice site of interest.

For small twists (away from the magic angle), the downfolding renormalizes and reduces the velocity $v(\theta)$ of the Dirac cones (see Figs.~\ref{fig:tbg_ldos_manytheta} and~\ref{fig:tbg_ldos}), but the low energy excitations remain semimetallic, as exemplified by the $\sim |E|/v(\theta)^2$ low energy behavior of the density of states. As the twist angle approaches the magic angle, hard gaps open at finite energy, forming an isolated miniband and a further reduced velocity, as shown in Fig.~\ref{fig:tbg_ldos_manytheta}. At the magic angle, the velocity vanishes and the miniband becomes nearly flat, resulting in a large density of states near zero energy.

\begin{figure}
\includegraphics[width=\linewidth]{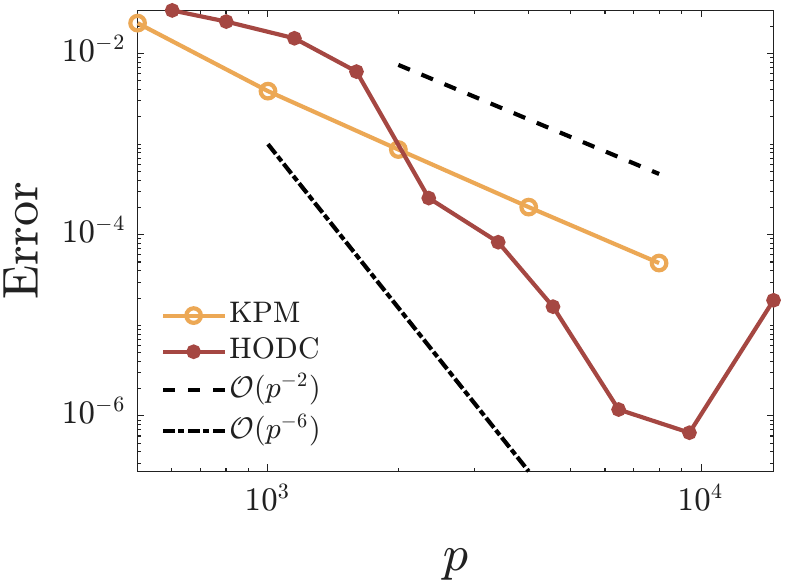} 
\caption{Pointwise relative error of Jackson KPM and HODC with $m = 6$ for the TBG example with $\theta = 6^\circ$ and $R_N = 3200$ at $\omega = -0.4$. For HODC, $\eta$ is varied, and the Chebyshev order $p$ is adaptively selected based on an error tolerance $\epsilon = 10^{-10}$.}
\label{fig:TBG_error_eta}
\end{figure}

We compute the LDOS using several different twist angles in Figs.~\ref{fig:tbg_ldos_manytheta} and \ref{fig:tbg_ldos}. In Fig.~\ref{fig:tbg_ldos}, for $\theta = 6^\circ$, we observe the emergence of the miniband structure as the broadening $\eta$ is decreased. For this example, in Fig.~\ref{fig:TBG_error_eta}, we plot the convergence of Jackson KPM and HODC with respect to $p$ at a fixed energy $E=-0.4$, which lies in the miniband. Here, the error is approximated by a comparison of results with increasing $p$. We observe the typical $\OO{p^{-2}}$ convergence of Jackson KPM, and approximate $\OO{p^{-6}}$ convergence of HODC with $m=6$. This example requires a significantly larger expansion order than the single-layer graphene example, due to the sophisticated structure of the spectrum. At the smooth point considered, for sufficiently large expansion order, we find that HODC reachs significantly higher accuracies than KPM.

\section{Conclusion}

We introduced the HODC method to compute the local density of states, which achieves high-order convergence at smooth points using a rational approximation of the $\delta$-function. For $m$th-order approximations, we obtain $\OO{\eta^m}$ convergence, in contrast with the $\OO{\eta^2}$ convergence of the standard KPM with Jackson regularization. Like KPM, the main computational step of HODC is the evaluation of the Chebyshev moments $\langle \bm r| T_n(H_N)|\bm r\rangle$ using the Chebyshev three-term recurrence. Our results show significantly higher attainable accuracy than KPM for the same computational effort.
We note that high accuracy is particularly useful within self-consistent calculations, such as mean field calculations, since self-consistent tolerances can then be achieved automatically.

Our method can be straightforwardly generalized to compute the global density of states using a stochastic estimation of the trace $\sum_{\bm r}\langle \bm r| T_n(H)|\bm r\rangle$. It can also be used to compute the imaginary part of the single particle Green's function $G(\bm r,\bm r',E)$ by considering $\langle \bm r| T_n(H)|\bm r' \rangle$. We expect that these ideas can also be extended to transport quantities and higher order response functions~\cite{KPM,joao2019basis,PhysRevB.110.014201,kubocomp20}, using multidimensional Chebshev expansions. We also expect our method can be applied to other KPM-based schemes, such as KPM + matrix product states~\cite{PhysRevB.83.195115}, and KPM + numerical renormalization group~\cite{PhysRevB.106.165123}.

\acknowledgments
We thank Olivier Parcollet, Miles Stoudenmire, and Angkun Wu for helpful discussions.
DM's research was supported by the National Science Foundation Grant No. 2422471. ML's research was partially supported by Simons Targeted Grant Award No. 896630.  ML's research was also supported by a visit to the Flatiron Institute's Center for Computational Quantum Physics while a portion of this research was carried out.   JHP has been supported in part by the US Department of Energy Office of Science, Basic Energy Sciences, under Award No.~DE-SC0026023 (work involved  conceptualization and construction of the project, as well as analysis and presentation of results).
The Flatiron Institute is a division of the Simons Foundation.

\appendix

\section{Convergence of the kernel polynomial method with Jackson regularization}
\label{appendix:jackson}

We verify the $\OO{p^{-2}}$ convergence of Jackson-regularized KPM, assuming $\rho_{\bf r}(E)$ is twice continuously differentiable~\cite{jackson}. We denote the Jackson kernel of order $p$ by $k_p^J(E,E')$, which is given by \eqref{eq:kpm_kernel} with regularization weights \eqref{eq:jackson_wgts}.
The Jackson local density approximation is
\begin{equation} \label{eq:ldoschebx}
    \rho_{\bm r}^{(p,N)}(E)  
    = \frac{1}{\pi\sqrt{1-E^2}}\sum_{n=0}^p g^p_n\mu_n T_n(E).
\end{equation}
Letting $\tilde \rho_{\bf r}(E) = \pi\sqrt{1-E^2}\rho_{\bf r}(E)$, Taylor's theorem gives
\begin{equation*}
\begin{split}
    &\int_{-1}^1 k_p^J(E,E')\rho_{\bf r}(E')\;dE' = \int_{-1}^1 \frac{k_p^J(E,E')}{\pi\sqrt{1-E'^2}}\tilde \rho_{\bf r}(E')\;dE'\\
    &= \int_{-1}^1 \frac{k^J_p(E,E')}{\pi\sqrt{1-E'^2}}\bigl[\tilde\rho_{\bf r}(E)+(E'-E) \tilde\rho_{\bf r}'(E)\\
    &\hspace{1cm}+\frac{1}{2}(E'-E)^2\tilde\rho_{\bf r}''(\tilde E)\bigr] dE',
\end{split}
\end{equation*} 
where $\tilde E(E') \in [E,E']$. For the first term,
\[
\int_{-1}^1 \frac{k^J_p(E,E')}{\pi\sqrt{1-E'^2}}\tilde\rho_{\bf r}(E)dE' = g_0^p \frac{\tilde \rho_{\bf r}(E)}{\pi\sqrt{1-E^2}} = \rho_{\bf r}(E).
\]
For the second term,
\[
 \int_{-1}^1 \frac{k_p^J(E,E')}{\pi\sqrt{1-E'^2}} (E'-E)\tilde \rho_{\bf r}'(E)dE' = (g_1^p-g_0^p)\frac{E \, \tilde \rho_{\bf r}'(E)}{\pi\sqrt{1-E^2}},
\]
which scales as $\OO{p^{-2}}$ since $g_1^p-g_0^p = \cos \frac{\pi}{p+1}-1$.
For the third term, we have
\begin{equation*}
    \begin{split}
        \biggl|\int_{-1}^1 &\frac{k_p^J(E,E')}{\pi\sqrt{1-E'^2}}(E'-E)^2\tilde \rho_{\bf r}''(\tilde E)\, dE' \biggr| \\
        &\leq \sup_{\tilde E \in [-1,1]} |\rho_{\bf r}''(\tilde E)| \; \int_{-1}^1 \frac{k_p^J(E,E')}{\pi\sqrt{1-E'^2}}(E'-E)^2dE'
    \end{split}
\end{equation*}
by the positivity of the kernel \cite{KPM}. Evaluating the integral on the right-hand side yields
\begin{equation*}
    \begin{split}
        \int_{-1}^1 &\frac{k_p^J(E,E')}{\pi\sqrt{1-E'^2}} (E'-E)^2 dE' \\
        &= \frac{\frac{g_0^p+g_2^p}{2}T_2(E) + g_0^p -2g_1^p T_1^2(E)}{\pi\sqrt{1-E^2}} \\
        &= \frac{(1 - g_2^p)(2E^2-1) + 1 - 2 E^2 \cos\frac{\pi}{p+1}}{\pi\sqrt{1-E^2}}.
    \end{split}
\end{equation*}

However since $g_2^p = 1 + \OO{p^{-2}}$, the error term scales as $\OO{p^{-2}}$, and
\begin{equation}
    \int k_p^J(E,E')\rho_{\bf r}(E')\;dE' = \rho_{\bf r}(E) + \OO{p^{-2}}.
\end{equation}

\section{High-order kernels}
\label{sec:appendix_HOK}

Here, we briefly review the definition, convergence properties, and construction of high-order rational regularization kernels. Following Ref.~\onlinecite[Def.~5.1]{colbrook21}, we say that an integrable kernel $K:\mathbb{R}\rightarrow\mathbb{R}$ \textit{is an $m$th-order kernel} if (i) it is normalized so that $\int_\mathbb{R} K(E')\,dE' = 1$, (ii) there is a constant $C_K>0$ such that $|K(E)|\leq C_K/(1+|E|)^{m+1}$ for all $E\in\mathbb{R}$, and (iii) it has vanishing moments $\int_\mathbb{R} K(E')E'^j\,dE' = 0$ for $j=1,\ldots,m-1$. These three criteria ensure that the shifted and dilated regularization kernel $K_\eta(E,E') = \eta^{-1}K(\eta^{-1}(E-E'))$ acts as a high-order approximation (in the weak sense) to $\delta(E-E')$ for small $\eta$. The following \textit{pointwise} rates of convergence for a regularized density are adapted from Ref.~\onlinecite[Thm.~5.2]{colbrook21}.
\begin{thm}
Let $\mu$ be a probability measure on the real line that is absolutely continuous on an interval $I=(E-\delta,E+\delta)$ for some fixed $E$ and $\delta>0$. Denote the Radon--Nikodym derivative of the absolutely continuous component of $\mu$ by $\rho$, and suppose that $\rho$ is $n$-times continuously differentiable on $I$ for some integer $n\geq 1$. If $K$ is an $m$th-order kernel, then the pointwise regularization error as $\eta\rightarrow 0^+$ satisfies
\begin{multline*}
|\rho(E)-\int_\mathbb{R} K_\eta(E,E')\,d\mu(E')| \\ \leq
\begin{cases}
    \OO{\eta^n}, &n<m, \\
    \OO{\eta^m\log \eta^{-1}}, &m\geq n.
\end{cases}
\end{multline*}
\end{thm}
\noindent
Note that the constants in the $\OO{\eta^m}$ and $\OO{\eta^n}$ terms depend on $\delta$, the distance to the nearest singularity of $\rho$, so that the high-order pointwise convergence deteriorates near the singularity.

To construct the rational $m$th-order regularization kernel in Eqn.~(\ref{eq:Keta}) we must ensure that $K_1(E,0)$ (i.e., the kernel before shifting and dilating) satisfies (i)-(iii). Given any distinct set of $m$ poles $z_1,\ldots,z_m$ in the upper half plane ($\Im z_k >0$ for $k=1,\ldots,m$), $K_1(E,0)$  satisfies (i)-(iii) if the weights $w_1,\ldots,w_m$ are chosen to satisfy the Vandermonde system~\cite{colbrook21}
$$
\begin{bmatrix}
    1 & 1 & \cdots & 1 \\
    z_1 & z_2 & \cdots & z_m \\
    \vdots & & & \vdots \\
    z_1^{m-1} & z_2^{m-1} & \cdots & z_m^{m-1}
\end{bmatrix}
\begin{bmatrix}
    w_1 \\ w_2 \\ \vdots \\ w_m
\end{bmatrix}
=
\begin{bmatrix}
    1 \\ 0 \\ \vdots \\ 0
\end{bmatrix}.
$$
For modest values of $m$ (say, $1\leq m\leq 8$), the weights can be computed numerically in standard IEEE double precision arithmetic. For larger values of $m$, the weights can be computed in extended precision and tabulated to mitigate numerical errors associated with solving the ill-conditioned Vandermonde system.

\bibliography{main}

\end{document}